\documentclass[3p]{elsarticle}

\usepackage{amssymb,amsmath}
\usepackage{bm}
\usepackage{graphicx}
\usepackage{tabularx}
\usepackage{comment}
\usepackage{fancyvrb}

\includecomment{pdffigure}

\DefineVerbatimEnvironment{code}{Verbatim}{fontsize=\small}
\newcommand{\openone}[0]{\leavevmode\hbox{\small1\normalsize\kern-.33em1}}

\biboptions{sort&compress}

\begin{document}

\begin{frontmatter}


\title{Plasmonics simulations with the MNPBEM toolbox: Consideration of substrates and layer structures}

\author{J\"urgen Waxenegger}
\author{Andreas Tr\"ugler}
\author{Ulrich Hohenester}
\ead{ulrich.hohenester@uni-graz.at}
\ead[url]{http://physik.uni-graz.at/~uxh}
\address{Institute of Physics, University of Graz,
  Universit\"atsplatz 5, 8010 Graz, Austria}



\begin{abstract}

Within the MNPBEM toolbox, developed for the simulation of plasmonic nanoparticles using a boundary element method approach, we show how to include substrate and layer structure effects.  We develop the methodology for solving Maxwell's equations using scalar and vector potentials within the inhomogeneous dielectric environment of a layer structure.  We show that the implementation of our approach allows for fast and efficient simulations of plasmonic nanoparticles situated on top of substrates or embedded in layer structures.  The new toolbox provides a number of demo files which can also be used as templates for other simulations.

\end{abstract}


\end{frontmatter}

\section*{Program summary}

\noindent{\sl Program title:} \texttt{MNPBEM} toolbox\\
\noindent{\sl Programming language:} Matlab 7.13.0 (R2011b)\\
\noindent{\sl Computer:} Any which supports Matlab 7.13.0 (R2011b)\\
\noindent{\sl Operating system:} Any which supports Matlab 7.13.0 (R2011b)\\
\noindent{\sl RAM required to execute with typical data:} $\ge 4$ GByte\\
\noindent{\sl Has the code been vectorised or parallelized?:} yes\\
\noindent{\sl Keywords:} Plasmonics, boundary element method, substrates and layer structures\\
\noindent{\sl CPC Library Classification:} Optics\\
\noindent{\sl External routines/libraries used:} no\\
\noindent{\sl Nature of problem:} Simulation of plasmonic nanoparticles placed on substrates or within layer structures\\
\noindent{\sl Solution method:} Boundary element method using electromagnetic potentials\\
\noindent{\sl Running time:} Depending on surface discretization between seconds and hours\\

\section{Introduction}\label{sec:intro}

Plasmonics provides an ideal tool for light confinement at the nanoscale~\cite{maier:07,atwater:07,schuller:10,stockman:11}.  A light wave excites coherent charge oscillations at the interface between metallic nanoparticles and a surrounding medium, so-called \textit{surface plasmons}, which come together with strongly localized, evanescent fields.  In a reversed process, quantum emitters such as molecules or quantum dots, located in the vicinity of plasmonic nanoparticles, couple to the evanescent fields and thereby use the nanoparticles as nano antennas to emit radiation more efficiently \cite{curto:10}.  Possible plasmonics applications range from photovoltaics, over sensorics and metamaterials, to optical and quantum-optical information processing \cite{halas:10}.

The simulation of plasmonic nanoparticles deals with the solution of Maxwell's equations.  For this reason, most of the simulation software builds on general Maxwell solvers such as the dyadic Green function technique \cite{chew:95,martin:95}, the finite difference time domain (FDTD) \cite{yee:66,ward:00,taflove:05,oskooi:10}, the discontinuous Galerkin time-domain (DGTD) \cite{niegemann:09}, or the discrete dipole approximation (DDA) \cite{draine:88,draine:94,loke:11} methods.  Another computational scheme is the boundary element method (BEM) \cite{garcia:02,hohenester.prb:05,myroshnychenko:08b,kern:09}, which is based on the assumption that the plasmonic nanoparticles consist of materials with homogeneous and isotropic dielectric functions separated by abrupt interfaces.

In the last couple of years we have developed a Matlab toolbox MNPBEM for the simulation of plasmonic nanoparticles using the BEM approach \cite{hohenester.cpc:12,hohenester.cpc:14b}.  This toolbox has been successfully employed by us and other groups.  It includes plane wave excitation and the computation of scattering, extinction, and absorption cross sections, as well as dipole excitations together with the computation of total and radiative scattering rates.  This allows one to compute the dyadic Green function or the photonic local density of states.  Additionally, we provide classes for the simulation of electron energy loss spectroscopy (EELS) of plasmonic nanoparticles \cite{hohenester.cpc:14b}.

In this paper we discuss a new version of the MNPBEM toolbox which allows for the simulation of plasmonic nanoparticles located in layer structures.  Indeed, many experiments are performed with nanoparticles situated on top of a substrate or embedded inside a layer structure, which calls for the possibility to perform corresponding simulations.  Unfortunately, for the potential-based BEM approach of Garc\'\i a de Abajo and coworkers \cite{garcia:02,myroshnychenko:08b,garcia:10} it has been unclear whether and how layer effects can be included.  In this paper we develop the methodology for BEM simulations including layer effects, and present a fast and flexible implementation within the MNPBEM toolbox.

\begin{figure}
\includegraphics[width=\columnwidth]{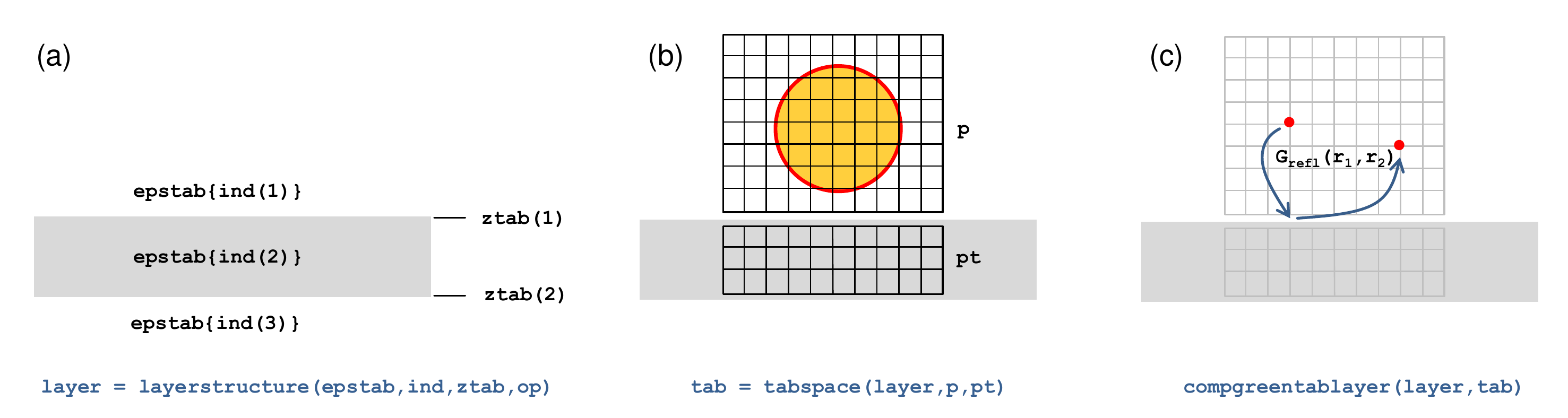}
\caption{Schematics of MNPBEM simulations including layer structures.  (a) We first set up a \texttt{layerstructure} by providing a table of dielectric functions, a pointer \texttt{ind} to this table for the layer materials in descending order, and the $z$-values of the layer interfaces.  (b) Next, we define a grid for the tabulation of the reflected Green functions.  With the command \texttt{tabspace(layer,p,pt)} we automatically generate the grids for a \texttt{comparticle} object \texttt{p} and an additional \texttt{compoint} object \texttt{pt}.  The latter points allow us to compute electromagnetic fields within the grid ranges.  (c) We finally set up an object \texttt{compgreentablayer} for the tabulated Green functions.  Once the layer structure is defined and the table of Green functions is pre-computed, BEM simulations can be performed as previously described for the MNPBEM toolbox without layer structures \cite{hohenester.cpc:12}.}\label{fig:layerstructure}
\end{figure}

To speed up the simulations, we compute at the beginning of each simulation a table of reflected Green functions which is then used for interpolation.  To include layer structure effects in plasmonics simulations, one must only define a few additional things in comparison to MNPBEM simulations of nanoparticles embedded in a homogeneous dielectric background (see Fig.~\ref{fig:layerstructure}):

\begin{enumerate}

\item First, one defines the layer structure.  The user must provide a table of dielectric functions, an index array that points to the materials of the different layers, and the positions of the layer interfaces.

\item With this layer structure one defines a grid on which the tabulated Green functions are computed.  The toolbox provides functions for setting up these grids either automatically (as we recommend) or manually.

\item In the next step one sets up a table for reflected Green functions and computes them for various wavelengths.  The underlying computation requires the evaluation of Sommerfeld-type integrals \cite{chew:95,paulus:00}, and is often rather slow.

\item Once the layer structure and the tabulated Green functions are computed, one can run the BEM simulations as previously described \cite{hohenester.cpc:14b}.

\end{enumerate}

We have organized this paper as follows.  In Sec.~\ref{sec:start} we discuss how to install the toolbox and give a few examples for plasmonics simulations with layer structures.  The methodology underlying our approach is presented in Sec.~\ref{sec:theory} and details about our implementation are given in Sec.~\ref{sec:implementation}.  In Sec.~\ref{sec:testing} we present a number or representative examples and compare our toolbox implementation with other approaches.  Finally, we summarize our approach in Sec.~\ref{sec:summary}.


\section{Getting started}\label{sec:start}

\subsection{Installation of the toolbox}

To install the toolbox, one must simply add the path of the main directory \texttt{mnpbemdir} of the \texttt{MNPBEM} toolbox as well as the paths of all subdirectories to the Matlab search path.  This can be done, for instance, through
\begin{code}
addpath(genpath(mnpbemdir));
\end{code}
To set up the help pages, one must once change to the main directory of the MNPBEM toolbox and run the program \texttt{makemnpbemhelp}
\begin{code}
>> cd mnpbemdir;
>> makemnpbemhelp;
\end{code}
Once this is done, the help pages, which provide detailed information about the toolbox, are available in the Matlab help browser.  The help pages can be found on the start page of the help browser under \textit{Supplemental Software}.  The toolbox is similar to our previously published versions~\cite{hohenester.cpc:12,hohenester.cpc:14b} but with a few modifications discussed in the following. 

\subsection{A simple example}

\begin{figure}
\centerline{\includegraphics[width=0.48\columnwidth]{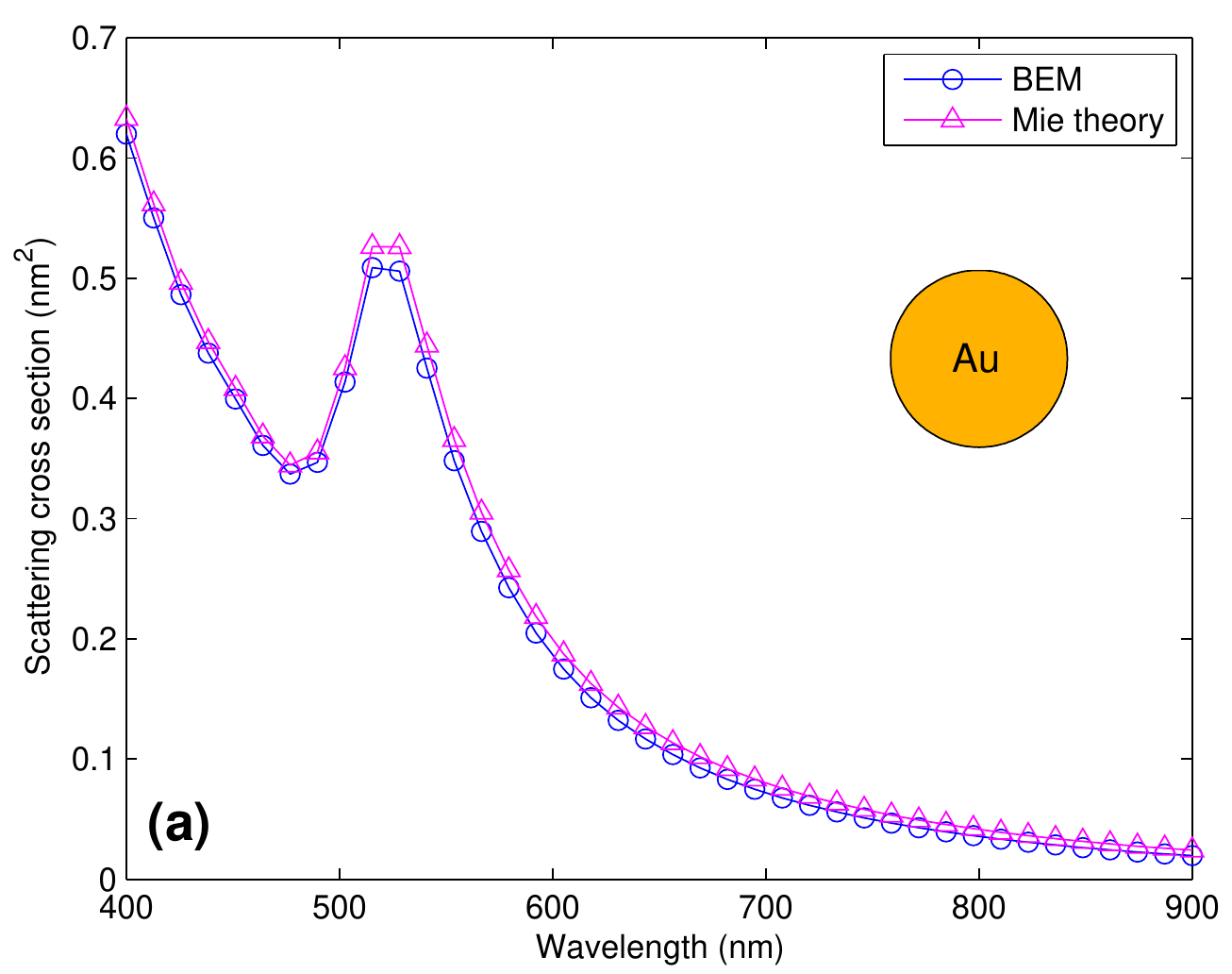}\quad
            \includegraphics[width=0.48\columnwidth]{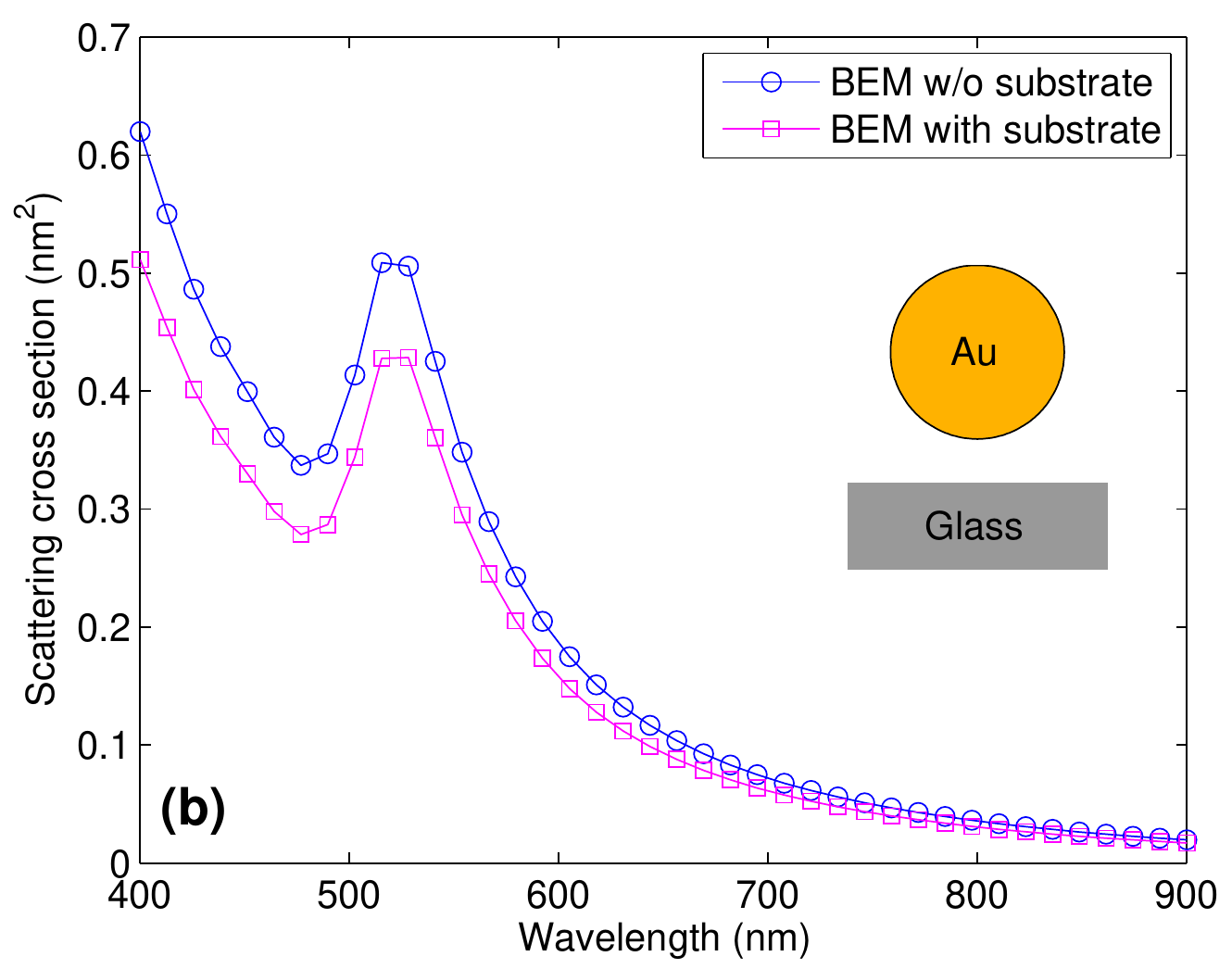}}
\caption{Planewave excitation of gold nanosphere, (a) without and (b) with layer structure.  The diameter of the sphere is 20 nm, the distance between substrate and sphere is 5 nm.  We assume plane wave excitation from above and use 144 vertices for the discretization of the sphere boundary.  The data of the gold dielectric functions are taken from optical experiments \cite{johnson:72}.  In panel (a) we compare our simulation results with Mie theory.}\label{fig:simple}
\end{figure}

\subsubsection{Scattering spectra of sphere without layer structure}

We first describe how to compute the scattering spectra for a sphere embedded in a homogeneous dielectric environment.  
\begin{code}
%  options for BEM simulation
op = bemoptions( 'sim', 'ret' );
%  table of dielectric functions
epstab = { epsconst( 1 ), epstable( 'gold.dat' ) };
%  initialize sphere
p = comparticle( epstab, { shift( trisphere( 144, 20 ), [ 0, 0, 15 ] }, [ 2, 1 ], 1, op );
\end{code}
In the first command line we set up an options structure and define that we want to perform a \textit{retarded} simulation using the full Maxwell equations.  In the second command line we define a table of dielectric functions, consisting of air ($\varepsilon=1$) and a dielectric function representative of gold~\cite{johnson:72}.  Finally, we set up a \verb|comparticle| object that defines the dielectric environment.  We define one boundary of spherical shape, with 144 vertices and a diameter of 20 nm, which is shifted into the upper half space (this has no effect for a sphere embedded in a homogeneous background, but will be important for layer structures).  The \verb|[2,1]| parameter indicates that the material \textit{inside} the boundary is \verb|epstab{2}|, and the \textit{outside} material is \verb|epstab{1}|.  We finally indicate that the nanosphere has a closed boundary and pass the options structure \verb|op| that controls the integration over boundary elements.  A more detailed description of the toolbox syntax has been given elsewhere \cite{hohenester.cpc:12,hohenester.cpc:14b} and can be also found in the help pages of the toolbox.

\begin{table}
\caption{Demo programs for simulations with layer structures provided by the MNPBEM toolbox.  We list the names of the programs, typical runtimes, and give brief explanations.  \texttt{stat} refers to demo files using the quasistatic approximation, with simulations using image charges, and \texttt{ret} to simulations of the full Maxwell equations.  The programs were tested on a standard PC (Intel i7--2600 CPU, 3.40 GHz, 8 GB RAM).}\label{table:examples}
{\small
\begin{tabularx}{\columnwidth}{lrX}
\hline\hline
Demo program & Runtime & Description \\
\hline
\texttt{demospecstat10.m} &  4 sec & Light scattering of metallic nanosphere above substrate \\
\texttt{demospecstat11.m} &  4 sec & Field enhancement for metallic sphere above substrate\\
\texttt{demospecstat12.m} &  8 sec & Light scattering of nanodisk above substrate\\
\texttt{demospecstat13.m} & 13 sec & Field enhancement for metallic disk above substrate\\
& & \\
\texttt{demospecret6.m} &  30 sec & Light scattering of metallic nanosphere above substrate   \\
\texttt{demospecret7.m} &  40 sec & Field enhancement of metallic nanosphere above substrate  \\
\texttt{demospecret8.m} &  1 min &  Scattering spectra for metallic nanodisk on substrate\\
\texttt{demospecret9.m} &  40 sec &  Scattering spectra for substrate using PARFOR loop\\
\texttt{demospecret10.m} & 47 sec &  Nearfield enhancement for metallic nanodisk on substrate \\
\texttt{demospecret11.m} & 2 min &  Scattering spectra for two nanospheres in layer\\
\texttt{demospecret12.m} & 1 min &   Nearfield enhancement for two nanospheres in layer\\
\texttt{demospecret13.m} & 8 min &  Spectra for metallic nanodisk approaching substrate \\
\texttt{demospecret14.m} & 15 min&  Spectra for metallic nanodisk on top of substrate\\
& & \\
\texttt{demodipstat5.m} & 8 sec & Lifetime reduction for dipole between sphere and layer \\
\texttt{demodipstat6.m} & 22 sec & Electric field for dipole between sphere and layer \\
\texttt{demodipstat7.m} & 36 sec & Photonic LDOS for nanodisk above layer \\
\texttt{demodipstat8.m} & 23 sec & Electric field for dipole close to nanodisk and layer \\
\texttt{demodipret8.m} & 24 sec & Lifetime reduction for dipole between sphere and layer \\
\texttt{demodipret9.m} & 50 sec & Electric field for dipole between sphere and layer  \\
\texttt{demodipret10.m} & 7 min & Photonic LDOS for nanodisk above layer  \\
\texttt{demodipret11.m} & 32 sec & Electric field for dipole close to nanodisk and layer \\
\hline
\hline
\end{tabularx}}
\end{table}

Once we have defined the dielectric environment, we can compute the scattering cross sections for different wavelengths \verb|enei| and plot them
\begin{code}
%  set up BEM solver
bem = bemsolver( p, op );
%  plane wave excitation
exc = planewave( [ 1, 0, 0 ], [ 0, 0, -1 ], op );
%  light wavelength in vacuum
enei = linspace( 400, 900, 40 );
%  allocate scattering cross section
sca = zeros( size( enei ) );

%  loop over wavelengths
for ien = 1 : length( enei )
  %  surface charge
  sig = bem \ exc( p, enei( ien ) );
  %  scattering cross section
  sca( ien, : ) = exc.sca( sig );
end

%  final plot
plot( enei, sca, 'o-'  );  hold on;

xlabel( 'Wavelength (nm)' );
ylabel( 'Scattering cross section (nm^2)' );
\end{code}
Figure~\ref{fig:simple}(a) shows the scattering spectra together with the corresponding Mie solutions.  We note that in comparison to the previous versions of the toolbox, we have now introduced wrapper functions \verb|bemsolver| and \verb|planewave| that select from the options structure the appropriate BEM solvers and excitation classes.

\subsubsection{Scattering spectra of sphere with layer structure}

The above example has to be only slightly modified for a simulation where the nanosphere is located above a substrate.  First we set up the layer structure
\begin{code}
%  table of dielectric functions
epstab = { epsconst( 1 ), epstable( 'gold.dat' ), epsconst( 2.25 ) };
%  set up layer structure
layer = layerstructure( epstab, [ 1, 3 ], 0 );
%  options for BEM simulation
op = bemoptions( 'sim', 'ret', 'layer', layer );
\end{code}
In the first command line we add an additional dielectric constant for glass to \verb|epstab|.  Next, we define a layer structure with the table of dielectric functions.  The second index argument \verb|[1,3]| indicates that the layer structure consists of two dielectric materials, the upper one is \verb|epstab{1}| and the lower one is \verb|epstab{3}|.  Finally, with the third input argument the layer interface is set to $z=0$.  Once the layer structure is defined, we must add it to the options structure.

An essential ingredient of BEM simulations with layer structures is the computation of reflected Green functions.  Within the MNPBEM toolbox, we compute at the beginning of the simulation a table of reflected Green functions, which is then added to the options structure
\begin{code}
%  automatic grid for tabulation
tab = tabspace( layer, p );
%  initialize Green function table
greentab = compgreentablayer( layer, tab );
%  precompute Green function table for wavelength array ENEI
greentab = set( greentab, enei, op );
%  add tabulated Green functions to options structure
op.greentab = greentab;
\end{code}
The meaning of the various command lines will be discussed in depth in the following sections and in the help pages of the toolbox.  In short, we first set up a grid \verb|tab| for the tabulated Green function.  Next, we set up a \verb|compgreentablayer| object that holds the tabulated Green functions, and compute them through the \verb|set| command.  Once the reflected Green function table is computed, we add it to the options structure.

The remaining simulation is identical to the one without layer structure.  Figure~\ref{fig:simple}(b) compares the scattering spectra without and with substrate.

The MNPBEM toolbox comes together with a number of demo files which are briefly described in Table~\ref{table:examples}.  We recommend to work through these demo files and to use them as templates for further simulations. 


\section{Theory}\label{sec:theory}

\subsection{BEM equations without layer structure}

We start by briefly reviewing the BEM approach developed by Garc\'\i a de Abajo and coworkers \cite{garcia:02,myroshnychenko:08b,garcia:10}  We consider dielectric nanoparticles, described through local and isotropic dielectric functions $\varepsilon_j(\omega)$, which are separated by sharp boundaries $\partial V_j$.  Throughout, we set the magnetic permeability $\mu=1$ and consider Maxwell's equations in frequency space $\omega$ \cite{jackson:99} (we adopt a Gaussian unit system).

The basic ingredients of the BEM approach are the scalar and vector potentials $\phi(\bm r)$ and $\bm A(\bm r)$, which are related to the electromagnetic fields via 
\begin{equation}
  \bm E=ik\bm A-\nabla\phi\,,\quad\bm B=\nabla\times\bm A\,.
\end{equation}
Here $k=\omega/c$ and $c$ are the wavenumber and speed of light in vacuum, respectively.  The potentials are connected through the Lorentz gauge condition $\nabla\cdot\bm A=ik\varepsilon\phi$.  Within each medium, we introduce the Green function for the Helmholtz equation defined through
\begin{equation}\label{eq:greenret}
  \left(\nabla^2+k_j^2\right)G_j(\bm r,\bm r')=-4\pi\delta(\bm r-\bm r')\,,\quad
  G_j(\bm r,\bm r')=\frac{e^{ik_j|\bm r-\bm r'|}}{|\bm r-\bm r'|}\,,
\end{equation}
with $k_j=\sqrt{\varepsilon_j}k$ being the wavenumber in the medium $\bm r\in V_j$.  For an inhomogeneous dielectric environment, we then write down the solutions of Maxwell's equations in the \textit{ad-hoc} form \cite{garcia:02,garcia:10}
\begin{subequations}\label{eq:adhoc}
\begin{eqnarray}
  \phi_j(\bm r)&=&\phi_j^e(\bm r)+
    \oint_{\partial V_j} G_j(\bm r,\bm s)\sigma_j(\bm s)\,da\label{eq:adhocphi}\\
  \bm A_j(\bm r)&=&\bm A_j^e(\bm r)+
    \oint_{\partial V_j} G_j(\bm r,\bm s)\bm h_j(\bm s)\,da\,, \label{eq:adhoca}
\end{eqnarray}
\end{subequations}
where $\phi_j^e$ and $\bm A_j^e$ are the scalar and vector potentials characterizing the external perturbation (e.g., plane wave or oscillating dipole) within a given medium $j$. Owing to Eq.~\eqref{eq:greenret}, these expressions fulfill the Helmholtz equations everywhere except at the particle boundaries.  $\sigma_j$ and $\bm h_j$ are surface charge and current distributions, which are chosen such that the boundary conditions of Maxwell's equations at the interfaces between regions of different permittivities $\varepsilon_j$ hold. 

In what follows, we introduce in accordance to Refs.~\cite{garcia:02,myroshnychenko:08b,garcia:10} matrix notations of the form $G\sigma$ instead of the integration given in Eq.~\eqref{eq:adhocphi}.  This also allows us to immediately change to a boundary element method (BEM) approach, where the boundary is split into elements of finite size suitable for a numerical implementation.  With $\sigma_1$ and $\bm h_1$ denoting the surface charges and currents at the particle \textit{insides}, and $\sigma_2$ and $\bm h_2$ the corresponding quantities at the particle \textit{outsides}, we obtain from the continuity of the scalar and vector potentials at the particle boundaries the expressions
\begin{equation}\label{eq:continuity1}
  G_1\sigma_1-G_2\sigma_2=\phi_2^e-\phi_1^e\,,\quad
  G_1\bm h_1-G_2\bm h_2=\bm A_2^e-\bm A_1^e\,.
\end{equation}
From the continuity of the Lorentz gauge condition and the dielectric displacement at the particle boundary, we get
\begin{subequations}\label{eq:continuity2}
\begin{eqnarray}
  H_1\bm h_1-H_2\bm h_2-ik\hat{\bm n}(\varepsilon_1G_1\sigma_1-\varepsilon_2G_2\sigma_2)&=&\bm\alpha\,,\quad\quad
  \bm\alpha=(\hat{\bm n}\cdot\nabla)(\bm A_2^e-\bm A_1^e)+ik\hat{\bm n}(\varepsilon_1\phi_1^e-\varepsilon_2\phi_2^e)
  \label{eq:contlorentz}\\
  \varepsilon_1H_1\sigma_1-\varepsilon_2H_2\sigma_2-ik\hat{\bm n}(\varepsilon_1G_1\bm h_1-\varepsilon_2G_2\bm h_2)&=&
  D^e\,,\quad D^e=\hat{\bm n}\cdot
  [\varepsilon_1(ik\bm A_1^e-\nabla\phi_1^e)-\varepsilon_2(ik\bm A_2^e-\nabla\phi_2^e)]\,,\qquad\label{eq:contdisp}
\end{eqnarray}
\end{subequations}
where $\hat{\bm n}$ is the outer surface normal of the boundary $\partial V$, and we have introduced the surface derivatives of the Green function $H_{1,2}=(\hat{\bm n}\cdot\nabla)G_{1,2}\pm 2\pi$.  Eqs.~\eqref{eq:continuity1} and \eqref{eq:continuity2} form a set of four coupled equations that can be solved within a boundary element method (BEM) approach in order to obtain the surface charges and currents, which provide a unique solution for the problem under study \cite{garcia:02,myroshnychenko:08b,garcia:10}.

\subsection{BEM equations with layer structure}

Suppose that we have a substrate or layer structure where the outer surface normals point in $z$-direction.  Inspection of Eq.~\eqref{eq:continuity2} reveals that in this case (i) the parallel component of $\bm h^\|$ does not couple with $h^\perp$ and $\sigma$, and (ii) $h^\perp$ and $\sigma$ become coupled through layer interactions.  This forms the basis of the BEM equations for layer structures.  First, we rewrite Eqs.~\eqref{eq:continuity1} and \eqref{eq:continuity2} for a layer structure.  Secondly, we express $\bm h^\|$ in terms of $h^\perp$ and $\sigma$.  Finally, we set up the coupled equations for $h^\perp$ and $\sigma$ and solve them within a boundary element method approach through matrix inversion.  Contrary to the BEM approach without layer structures, which only deals with matrices of order $N$ of the number of boundary elements, the BEM approach with layer structures deals with matrices of the order $2N$.

We consider a nanoparticle located in the dielectric environment of a layer structure and assume that all boundary elements connected to the layer structure are \textit{outer} elements (defined with respect to the surface normals and indicated with 2).  In the spirit of Ref.~\cite{garcia:02}, the potentials \textit{inside} the nanoparticle can still be expressed in the ad-hoc form of Eq.~\eqref{eq:adhoc}.  For the boundary elements \textit{outside} the nanoparticle, we (i) have to replace $G$ by the Green function for the layer structure and (ii) have to account for the fact that $h_2^\perp$ and $\sigma_2$ become coupled,
\begin{equation}
  \phi_2=\phi_2^e+G_2^{\sigma\sigma}\sigma_2+G_2^{\sigma h}h_2^\perp\,,\quad
  A_2^\perp=A_2^{e\perp}+G_2^{hh}h_2^\perp+G_2^{h\sigma}\sigma_2\,.
\end{equation}
We will show in Sec.~\ref{sec:refl} how to compute the various reflected Green functions.  Next, we re-derive the BEM equations of the previous section for layer structures.  The continuity of the potentials now becomes
\begin{subequations}\label{eq:continuity3}
\begin{eqnarray}
  G_1\sigma_1&=&G_2^{\sigma\sigma}\sigma_2+G_2^{\sigma h}h_2^\perp+\varphi\,,\qquad\,\, \varphi=\phi_2^e-\phi_1^e\\
  G_1\bm h_1^\| &=& G_2^\|\bm h_2^\|+\bm a^\| \\
  G_1h_1^\perp&=&G_2^{hh}h_2^\perp+G_2^{h\sigma}\sigma_2+a^\perp\,,\qquad \bm a=\bm A_2^e-\bm A_1^e\,.
\end{eqnarray}
\end{subequations}
The continuity of the Lorentz condition reads
\begin{subequations}\label{eq:continuity4}
\begin{eqnarray}
  H_1\bm h_1^\|-H_2\bm h_2^\|-ik\hat{\bm n}^\|\left(\varepsilon_1 G_1\sigma_1-
  \varepsilon_2 G_2^{\sigma\sigma}\sigma_2-\varepsilon_2 G_2^{\sigma h}h_2^\perp\right) &=& \bm\alpha^\|\\
  H_1 h_1^\perp- H_2 h_2^\perp-H_2^{h\sigma}\sigma_2-ik\hat{n}^\perp\left(\varepsilon_1 G_1\sigma_1-
  \varepsilon_2 G_2^{\sigma\sigma}\sigma_2-\varepsilon_2 G_2^{\sigma h}h_2^\perp\right) &=& \alpha^\perp\,,
\end{eqnarray}
\end{subequations}
and the continuity of the dielectric displacement becomes
\begin{equation}\label{eq:continuity5}
  \varepsilon_1 H_1\sigma_1-\varepsilon_2 H_2^{\sigma\sigma}\sigma_2-\varepsilon_2 H_2^{\sigma h}h_2^\perp-
  ik\hat{\bm n}^\|\cdot\bigl(\varepsilon_1 G_1\bm h_1^\|-\varepsilon_2 G_2^\|\bm h_2^\|\bigr)-
  ik\hat{n}^\perp\bigl(\varepsilon_1 G_1h_1^\perp-\varepsilon_2 G_2^{hh}h_2^\perp-
  \varepsilon_2 G_2^{h\sigma}\sigma_2\bigr)=D^e\,.
\end{equation}
The set of Eqs.~(\ref{eq:continuity3}--\ref{eq:continuity5}) now has to be solved to obtain the surface charges and currents.\footnote{In Ref.~\cite{garcia:02} the authors used $\varepsilon_2G_2=G_2\varepsilon_2$ in order to obtain the $L$ matrices in case of an arbitrary number of media.  Such exchange is possible because $G_2$ only connects points within the same medium.  For layer structures, this exchange is no longer allowed since the reflected Green functions can connect points in different layers and different media.  The generalization to arbitrary number of media is still possible for layer structures, but $\varepsilon$ has to be interpreted as a diagonal matrix and $G_2$ and $\varepsilon_2$ must not be exchanged.}  After some rearrangements, outlined in \ref{sec:working}, we arrive at the working equations for BEM with layer structures
\begin{subequations}\label{eq:working}
\begin{eqnarray}
  &&\left(\varepsilon_1\Sigma_1 G_2^{\sigma\sigma}-\varepsilon_2H_2^{\sigma\sigma}\right)\sigma_2+
  \left(\varepsilon_1\Sigma_1 G_2^{\sigma h}-\varepsilon_2 H_2^{\sigma h}\right)h_2^\perp
  -ik\hat{\bm n}^\|\cdot\Gamma\hat{\bm n}^\|(\varepsilon_1-\varepsilon_2)
  \left(G_2^{\sigma\sigma}\sigma_2+G_2^{\sigma h}h_2^\perp\right)-\nonumber\\
  &&\qquad ik\hat n^\perp(\varepsilon_1-\varepsilon_2)
  \left(G_2^{h\sigma}\sigma_2+G_2^{hh}h_2^\perp\right)=
  D^e-\varepsilon_1\Sigma_1\varphi+ik\hat{\bm n}\cdot\varepsilon_1\bm a+
  \hat{\bm n}^\|\cdot\Gamma\left(\bm\alpha^\|-\Sigma_1\bm a^\|+ik\hat{\bm n}^\|\varepsilon_1\varphi\right)
  \qquad\\
  && \left(\Sigma_1 G_2^{h\sigma}-H_2^{h\sigma}\right)\sigma_2+
  \left(\Sigma_1G_2^{hh}-H_2^{hh}\right)h_2^\perp-
  ik\hat n^\perp(\varepsilon_1-\varepsilon_2)\left(G_2^{\sigma\sigma}\sigma_2+G_2^{\sigma h}h_2^\perp\right)
  =\alpha^\perp-\Sigma_1 a^\perp+ik\hat n^\perp\varepsilon_1\varphi\,,\quad\nonumber\\
\end{eqnarray}
\end{subequations}
with $\Gamma=ik(\varepsilon_1-\varepsilon_2)(\Sigma_1-\Sigma_2^\|)^{-1}$.  The set of equations (\ref{eq:working}a,b) can be interpreted as a matrix equation for $(\sigma_2,h_2^\perp)$ which is solved through matrix inversion.  Once $\sigma_2$ and $h_2^\perp$ are obtained, we get $\bm h_2^\|$ through the solution of Eq.~\eqref{eq:app1} and $\sigma_1$, $\bm h_1$ through the solution of Eq.~\eqref{eq:continuity3}.  The working equations of Eq.~\eqref{eq:working} are somewhat more complicated than those of the original BEM approach \cite{garcia:02}, but still provide a numerically tractable approach.

\subsection{Reflected Green functions}\label{sec:refl}

The Green functions $G_2^\|$, $G_2^{\sigma\sigma}$, $G_2^{\sigma h}$, $G_2^{h \sigma}$, and $G_2^{hh}$ are essential ingredients of the BEM approach for layer structures.  They are computed similarly to related field-based approaches \cite{chew:95,paulus:00}.  Consider a boundary element with surface charges $\sigma$ and currents $\bm h$ within a layer structure.  $\sigma$ and $\bm h$ lead to potentials $\phi^e=G\sigma$ and $\bm A^e=G\bm h$ impinging at the interfaces of the layer structure.  In accordance to field-based approaches, we (i) expand the scalar and vector potentials originating from the source points in cylinder waves, (ii) compute the surface charges and currents at the interfaces by using the BEM equation (this step is similar to obtaining the Fresnel reflection and transmission coefficients), and (iii) finally compute the potentials at the observation points by integrating over all cylinder waves.

In step (i) we employ the usual Sommerfeld identity~\cite{chew:95}
\begin{equation}
  \frac{e^{ikr}}r=i\int_0^\infty \frac{k_\rho}{k_z}J_0(k_\rho\rho)e^{ik_zz}\,dk_\rho\,,
\end{equation}
with the wavevector $k$ decomposed into radial component $k_\rho$ and $z$-component $k_z=\sqrt{k^2-k_\rho^2}$.  $J_0$ is the Bessel function of order zero.  The wave $e^{ik_zz}$ impinging at the interface becomes reflected and transmitted, and in step (iii) we have to sum over all reflected and transmitted waves
\begin{equation}\label{eq:sommerfeld}
  i\int_0^\infty \frac{k_\rho}{k_z}J_0(k_\rho\rho)e^{ik_zz}A(k_\rho,k_z)\,dk_\rho\,,
\end{equation}
where $A(k_\rho,k_z)$ is a generalized reflection or transmission coefficient to be discussed below.  To evaluate the integral of Eq.~\eqref{eq:sommerfeld} we directly follow Ref.~\cite{paulus:00} and deform the integration path in the complex plane using the recipes given in this paper.

To compute the BEM reflection and transmission coefficients for the layer structure, we proceed as follows.  Consider a layer structure with interfaces at $z_\mu$.  We denote the medium \textit{above} the layer with $\mu$ and the medium \textit{below} the layer with $\mu+1$.  Thus, $\mu=1$ denotes the uppermost medium.  We assume that the outer surface normal points into the positive $z$-direction, and denote the surface charges and currents at the upper side of $z_\mu$ with $\sigma_2^\mu$ and $\bm h_2^\mu$, and at the lower side of $z_\mu$ with $\sigma_1^{\mu+1}$ and $\bm h_1^{\mu+1}$.  Additionally, we introduce the intralayer Green functions $G_0^\mu$ that connect points in layer $z_\mu$ and in medium $\mu$, and interlayer Green functions $G^\mu$ that connect points between different layers in medium $\mu$.  Let $\phi_{1,2}^\mu$ and $\bm A_{1,2}^\mu$ denote the external excitations described by scalar and vector potentials, respectively.

As the BEM equations (\ref{eq:continuity1},\ref{eq:continuity2}) decouple $\bm h^\|$ from $\sigma$, $h^\perp$, we can treat excitations $\bm A^\|$ and $\phi$, $A^\perp$ separately.  For parallel excitations $\bm A_{1,2}^\mu$, we obtain the following set of equations for the parallel surface currents $\bm h_{1,2}^\mu$
\begin{subequations}\label{eq:parallel}
\begin{eqnarray}
  G_0^{\mu+1}\bm h_1^{\mu+1}-G_0^\mu\bm h_2^\mu-G^\mu\bm h_1^\mu+G^{\mu+1}\bm h_2^{\mu+1}&=&
  \bm A_2^\mu-\bm A_1^{\mu+1}\\
  2\pi i\left(\bm h_1^{\mu+1}+\bm h_2^\mu\right)-
  k_z^\mu G^\mu\bm h_1^\mu-k_z^{\mu+1}G^{\mu+1}\bm h_2^{\mu+1}&=&
   k_z^\mu\bm A_2^\mu+k_z^{\mu+1}\bm A_1^{\mu+1}\,,
\end{eqnarray}
\end{subequations}
which can be solved for each wavevector through matrix inversion.  For a perpendicular vector potential $A_{1,2}^\mu$ or a scalar potential $\phi_{1,2}^\mu$ the BEM equations become
\begin{subequations}\label{eq:perp}
\begin{eqnarray}
  &&G_0^{\mu+1}\sigma_1^{\mu+1}-G_0^\mu\sigma_2^\mu-G^\mu\sigma_1^\mu+G^{\mu+1}\sigma_2^{\mu+1}=
  \phi_2^\mu-\phi_1^{\mu+1} \\
  &&G_0^{\mu+1}h_1^{\mu+1}-G_0^\mu h_2^\mu-G^\mu h_1^\mu+G^{\mu+1}h_2^{\mu+1}=
  A_2^\mu-A_1^{\mu+1} \\
  && 2\pi i\left(\varepsilon_{\mu+1}\sigma_1^{\mu+1}+\varepsilon_\mu\sigma_2^\mu\right)+
  k\left(G_0^{\mu+1}\varepsilon_{\mu+1}h_1^{\mu+1}-G_0^\mu\varepsilon_\mu h_2^\mu\right)-
  \nonumber\\&&\qquad\qquad
  k_z^\mu\varepsilon_\mu G^\mu\sigma_1^\mu-k_z^{\mu+1}\varepsilon_{\mu+1}G^{\mu+1}\sigma_2^{\mu+1}
  -k\varepsilon_\mu G^\mu h_1^\mu+k \varepsilon_{\mu+1}G^{\mu+1}h_2^{\mu+1}=
  \nonumber\\&&\qquad\qquad
  k_z^\mu\varepsilon_\mu\phi_2^\mu+k_z^{\mu+1}\varepsilon_{\mu+1}\phi_1^{\mu+1}+
  k\varepsilon_\mu A_2^\mu-k\varepsilon_{\mu+1} A_1^{\mu+1}\\
  &&2\pi i\left(h_1^{\mu+1}+h_2^\mu\right)+
  k\left(G_0^{\mu+1}\varepsilon_{\mu+1}\sigma_1^{\mu+1}-G_0^\mu\varepsilon_\mu\sigma_2^\mu\right)-
  \nonumber\\&&\qquad\qquad
  k_z^\mu G^\mu h_1^\mu-k_z^{\mu+1}G^{\mu+1}h_2^{\mu+1}-
  k\varepsilon_\mu G^\mu\sigma_1^\mu+k\varepsilon_{\mu+1}G^{\mu+1}\sigma_2^{\mu+1}=
  \nonumber\\&&\qquad\qquad
  k_z^\mu A_2^\mu+k_z^{\mu+1}A_1^{\mu+1}+k\varepsilon_\mu\phi_2^\mu-k\varepsilon_{\mu+1}\phi_1^{\mu+1}\,.
\end{eqnarray}
\end{subequations}
Equations \eqref{eq:parallel} and \eqref{eq:perp} can be used to compute the reflected Green functions for layer structures.  For instance, to compute $G^{h\sigma}$ we consider an exciting scalar potential, produced by a surface charge $\sigma$ at the source point, and compute the perpendicular component of the vector potential produced by the induced surface current distribution $h_{1,2}^\mu$ at the observation point.  As the solutions of Eqs.~\eqref{eq:parallel} and \eqref{eq:perp} involve the inversion of matrices of low order (number of layer media multiplied with two or four), this computation is extremely fast.


\section{Implementation}\label{sec:implementation}

In this section we describe how simulations with layer structures have been implemented within the MNPBEM toolbox.

\subsection{Layer structure}

To set up a layer structure, one must provide a table of dielectric functions (the \textit{same} table \verb|epstab| as used in the initialization of \verb|comparticle| objects), the interface positions \verb|ztab|, and an index array \verb|ind| that points to the different media in descending order.  \verb|epstab{ind(1)}| is the dielectric function of the uppermost medium above \verb|ztab(1)|, \verb|epstab{ind(2)}| is the dielectric function for the next medium, and so forth.  With these arrays, we set up a layer structure with
\begin{code}
%  options for layer structure
op = layerstructure.options( 'rmin', 1e-2 );
%  set up layer structure
layer = layerstructure( epstab, ind, ztab, op );
\end{code}
\verb|op| is an options structure with the following fields:
\begin{description}

\item[] \texttt{ztol} (default value \verb|0.02|) Boundary elements located closer than \verb|ztol| to the layer interfaces are assumed to belong to the layer, see discussion below.

\item[] \texttt{rmin} (default value \verb|0.01|) Minimum radial distance for the evaluation of the reflected Green functions.

\item[] \texttt{zmin} (default value \verb|0.01|) Minimum $z$-value for the evaluation of the reflected Green functions.

\item[] \texttt{semi} (default value \verb|0.1|) Ratio of semi-axes $k_\rho^{\rm min}:k_\rho^{\rm max}$ for the integration in the complex plane, as described in Ref.~\cite{paulus:00}.

\item[] \texttt{ratio} (default value \verb|2|) Ratio $z:r$ which determines whether an integration with Bessel or Hankel functions is performed, as described in Ref.~\cite{paulus:00}.

\item[] \texttt{op} Additional options passed to the Matlab \texttt{ode45} integration routine for the evaluation of the Sommerfeld integrals in the complex plane.

\end{description}

The \verb|layerstructure| class has several functions for the evaluation of the reflected Green functions.  More details can be found by typing 
\begin{code}
>> doc layerstructure
\end{code}
at the Matlab prompt or by consulting the help pages of the toolbox.  Once the layer structure is defined, it must be added to the options structure
\begin{code}
%  add layer structure to options
op.layer = layer;
\end{code}

\subsection{Tabulated Green functions}

Quite generally, the evaluation of the reflected Green functions is rather time consuming and in many cases constitutes a serious bottleneck for BEM simulations.  In order to speed up the simulations, in the MNPBEM toolbox we first set up a table of reflected Green functions, using a suitable grid for the different radii $r$ and $z$-values, and then perform an interpolation.

The reflected Green functions depend on $G(r,z_1,z_2)$, where $r$ is the radial distance between the observation and source point, $z_1$ is the $z$-value of the observation point, and $z_2$ is the $z$-value of the source point.  For the uppermost or lowermost medium the Green functions only depend on $z_1+z_2$ \cite{chew:95}, which allows for an additional speedup.  Within our BEM approach, we additionally need the derivatives $F_r=\partial G/\partial r$ and $F_z=\partial G/\partial z_1$ for the evaluation of $H_2$, see Eq.~\eqref{eq:working}.  Instead of directly interpolating $G$, $F_r$, and $F_z$, we assume a functional dependence of the form
\begin{equation}\label{eq:interp}
  G(r,z_1,z_2)=\frac{g(r,z_1,z_2)}{\tilde r}\,,\quad
  F_r(r,z_1,z_2)=-\frac{f_r(r,z_1,z_2)\,r}{\tilde r^3}\,,\quad
  F_z(r,z_1,z_2)=-\frac{f_z(r,z_1,z_2)\,\tilde z}{\tilde r^3}\,,\quad
\end{equation}
where $\tilde z$ is the sum of the $z_1$ and $z_2$ distances to the respective closest layer interfaces, $\tilde r=\sqrt{r^2+\tilde z^2}$, and $g$, $f_r$, $f_z$ are tabulated values.  Eq.~\eqref{eq:interp} has the advantage that for small $r$ and $z$-values the functional shape is the same as for quasistatic Green functions using image charges \cite{jackson:99}, and we expect that for layer structures $g$, $f_r$, and $f_z$ in general have only a weak spatial dependence, as also confirmed by our test simulations.

To set up the table of reflected Green functions, we need to define a grid for the tabulated $r$, $z_1$, and $z_2$ values.  This can be either done automatically (as we recommend) or manually.  Through
\begin{code}
%  Green function mesh for BEM simulation with particle P
tab = tabspace( layer, p );
%  mesh for particle P, and compute fields at COMPOINT positions PT
tab = tabspace( layer, p, pt );
%  additionally provide number of radii and z-values NR and NZ
tab = tabspace( layer, p, 'nr', nr, 'nz', nz );
\end{code}
we set up an automatic grid with logarithmic spacings.  Alternatively, one can also set up a structure
\begin{code}
%  set up table manually 
tab = struct( 'r', rtab, 'z1', z1tab, 'z2', z2tab );
\end{code}
The minimal \verb|rtab| and \verb|ztab| values should be equal to the \verb|layer.rmin| and \verb|layer.zmin| values.  It is important that \verb|tab| only connects points within a single layer medium.  For points located in different media, one must provide an array of \verb|tab| values, as produced automatically with the calling sequence involving \verb|p| and \verb|pt| (see also help pages for a more detailed description).

\begin{figure}
\centerline{\includegraphics[width=0.5\columnwidth]{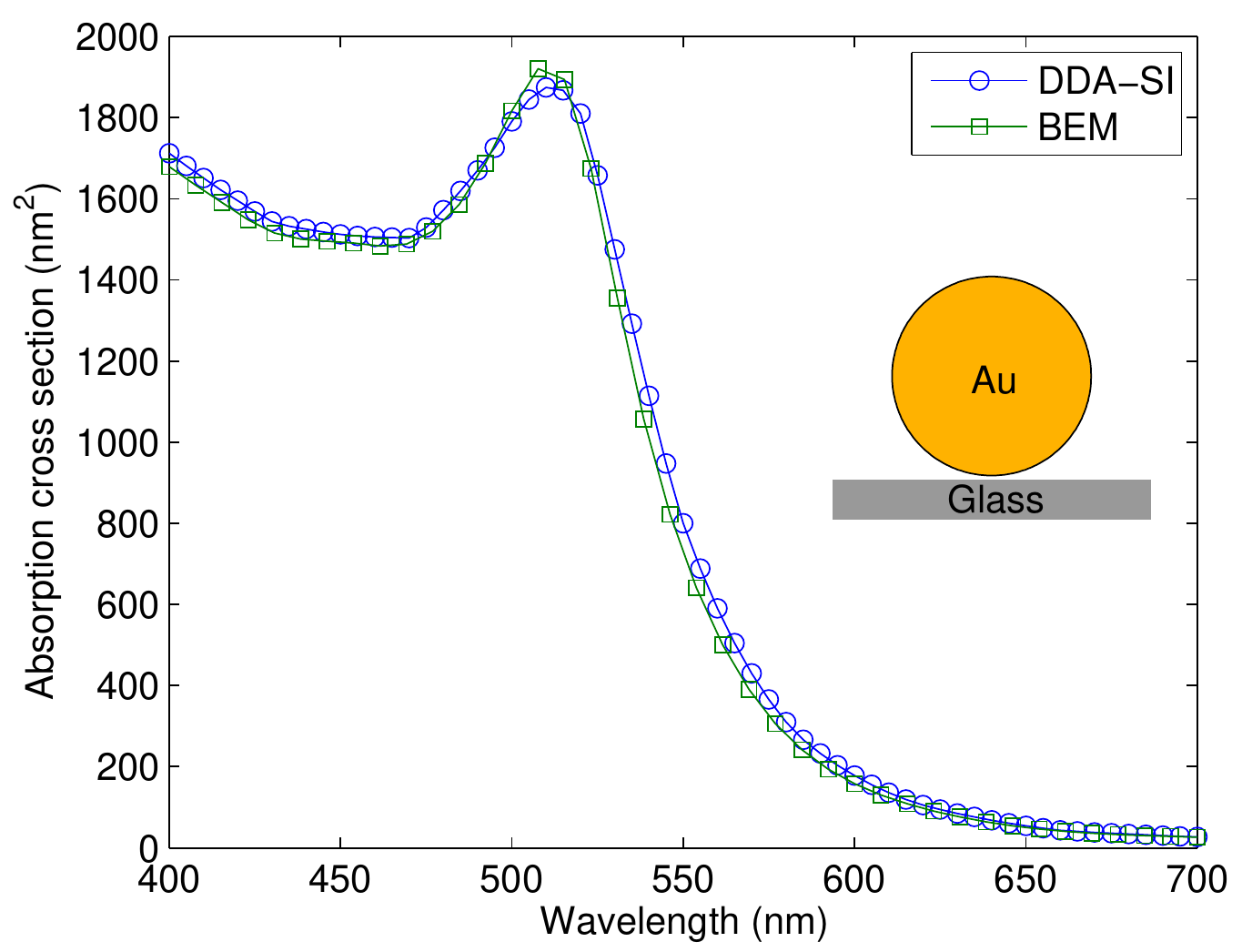}}
\caption{Comparison between DDA simulation for substrates (DDA--SI, \cite{loke:11}) and our BEM implementation.  We plot the absorption cross section for a gold nanosphere (50 nm diameter) located 1 nm above a glass substrate (refractive index $n=1.52$).  For the DDA--SI simulation we use 1472 dipoles, for the BEM simulation we use the sphere discretization \texttt{trisphere(625,50)} with 1246 boundary elements.  In \cite{loke:11} the authors additionally compared their results with FDTD simulations, finding perfect agreement for the entire wavelength regime.}\label{fig:dda}
\end{figure}

Once the grid is defined, one can precompute the Green function table and add it to the options structure
\begin{code}
%  initialize Green function table
greentab = compgreentablayer( layer, tab );
%  precompute Green function table for wavelength array ENEI
greentab = set( greentab, enei, op );
%  add tabulated Green functions to options structure
op.greentab = greentab;
\end{code}
Depending on the grid size and the layer structure, the \verb|set| call can be rather time-consuming.  For this reason, we recommend to compute the table only if it has not been computed before.  Quite generally, a \verb|greentab| table with proper grid size can be used for various simulations sharing the same layer structure.  One may thus consider saving it on the hard disk.  The toolbox has a \verb|layerstructure/ismember| function that checks whether a previously computed Green function table is compatible with other simulations.  We also provide a parallel version \verb|parset| for precomputation.  More details about these features can be found in the toolbox help pages.

\begin{figure}[t]
\centerline{\includegraphics[width=0.65\columnwidth]{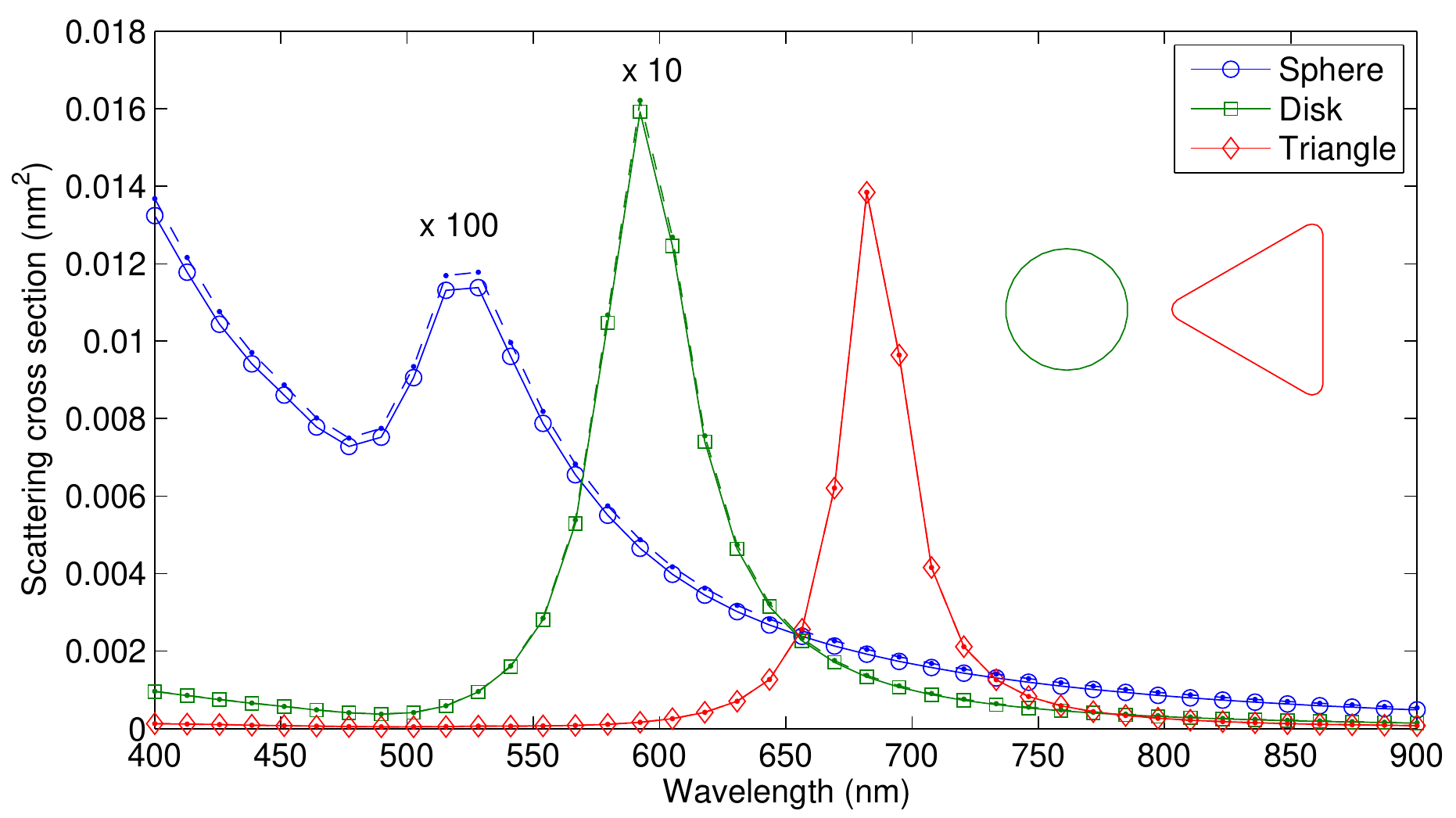}}
\caption{Comparison of scattering spectra for different gold nanoparticles, and for retarded (full lines) and quasistatic (dashed-dotted lines, almost indistinguishable) simulations.  For the quasistatic simulations, the substrate (glass) is modeled by the method of image charges.  The nanosphere is produced with \texttt{shift(trisphere(256,5),[0,0,3])} and the substrate interface is located at $z=0$.  For the nanodisk we use \texttt{tripolygon(poly,edge)} with \texttt{poly=polygon(30,'size',[6,6])} and \texttt{edge=edgeprofile(1,11,'min',0.5)}, and for the nanotriangle \texttt{poly=round(polygon(3,'size',[8,8*2/sqrt(3)]))}.  The gold dielectric function is taken from \cite{johnson:72}.  The cross sections of the sphere and disk are multiplied by factors of 100 and 10, respectively.  In the inset we show the contours for the disk and triangle, respectively.
}\label{fig:stat}
\end{figure}

\subsection{BEM simulations with layer structures}

Once the direct and reflected Green functions have been computed they can be directly used for the solution of the BEM equations \eqref{eq:working}.  There is one critical issue regarding boundary elements that are directly located on an interface.  Before pondering on such interface elements, we recall that in the normal BEM approach one has to be careful when computing the surface derivative of the Green function for diagonal elements~\cite{garcia:02}
\begin{equation}\label{eq:singular}
  \lim_{\bm r\to\bm s}(\hat{\bm n}\cdot\nabla_{\bm r})G(\bm r,\bm s')=
  (\hat{\bm n}\cdot\nabla_{\bm s})G(\bm s,\bm s')\pm 2\pi\delta(\bm s-\bm s')\,,
\end{equation}
where the sign of the singular term depends on whether $\bm r$ approaches $\bm s$ from the inside or outside.  Inspection of Eq.~\eqref{eq:interp} shows that a similar singular contribution shows up in the reflected Green function for elements belonging to an interface.  Within the toolbox, boundary elements of \verb|p| must be located either slightly above or below the interface.  If the centroid position \verb|p.pos| is closer than \verb|layer.ztol| to the interface, the program assumes that it belongs to the interface.  The surface derivative $F_z$ of the reflected Green function is now split into two contributions (note that $\hat{\bm n}$ points into the $z$-direction)
\begin{equation}\label{eq:singularlayer}
  F_z(r,z_1,z_2)=-f_0\frac{\tilde z}{\tilde r^3}-\bigl(f_z(r,z_1,z_2)-f_0\bigr)\frac{\tilde z}{\tilde r^3}\,,
\end{equation}
where $f_0=\lim_{r,\tilde z\to 0}f_z(r,\tilde z)$.  When approaching the boundary through $\lim_{\bm r\to\bm s}F_z(r,z_1,z_2)$, the first term gives a singular contribution $\pm 2\pi f_0\delta(\bm s-\bm s')$, similarly to Eq.~\eqref{eq:singular}, whereas the second term has a smooth $r$ and $z$ dependence and can be safely integrated over the boundary element.


\section{Testing the toolbox}\label{sec:testing}

\begin{figure}
\centerline{\includegraphics[width=0.48\columnwidth]{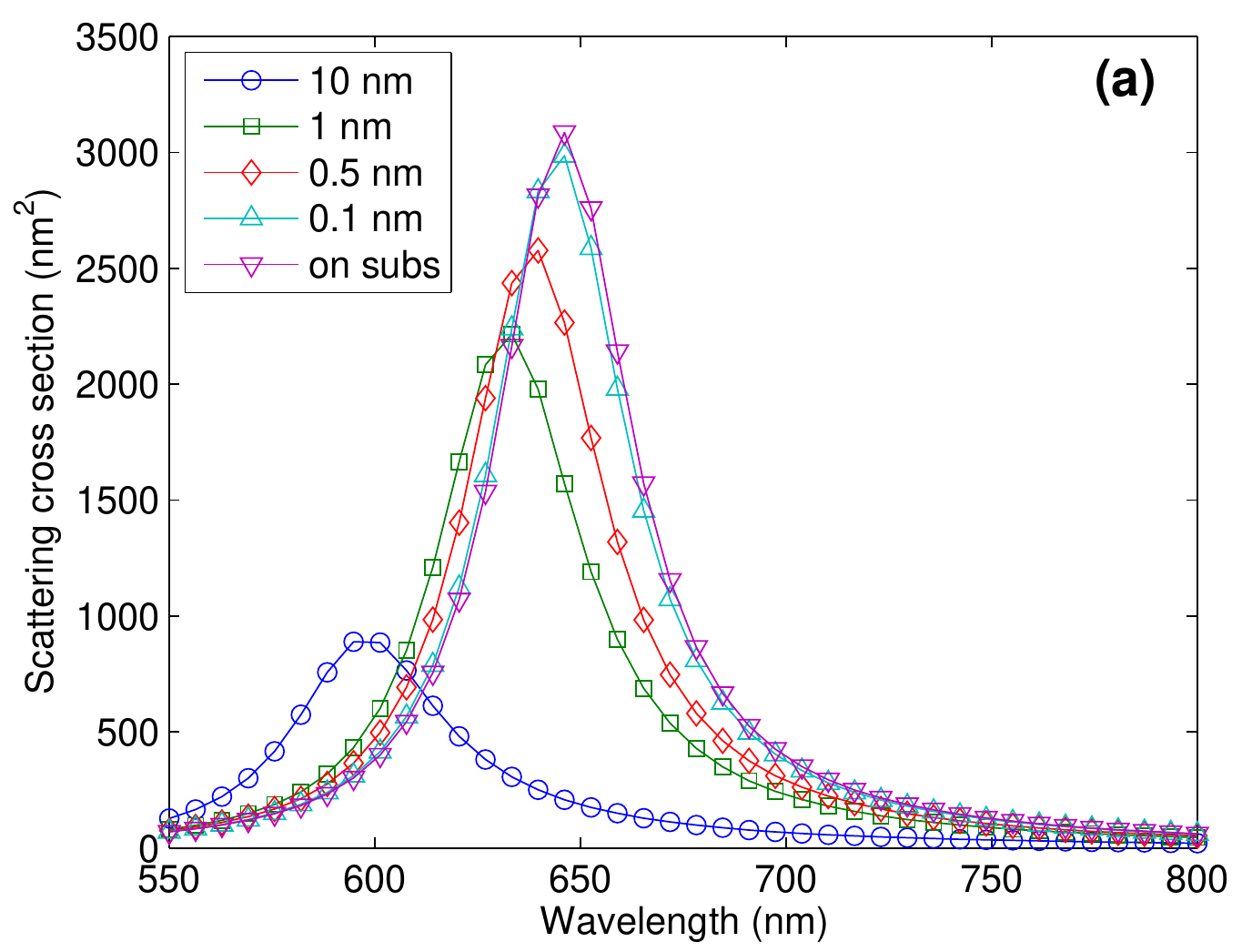}\quad
            \includegraphics[width=0.48\columnwidth]{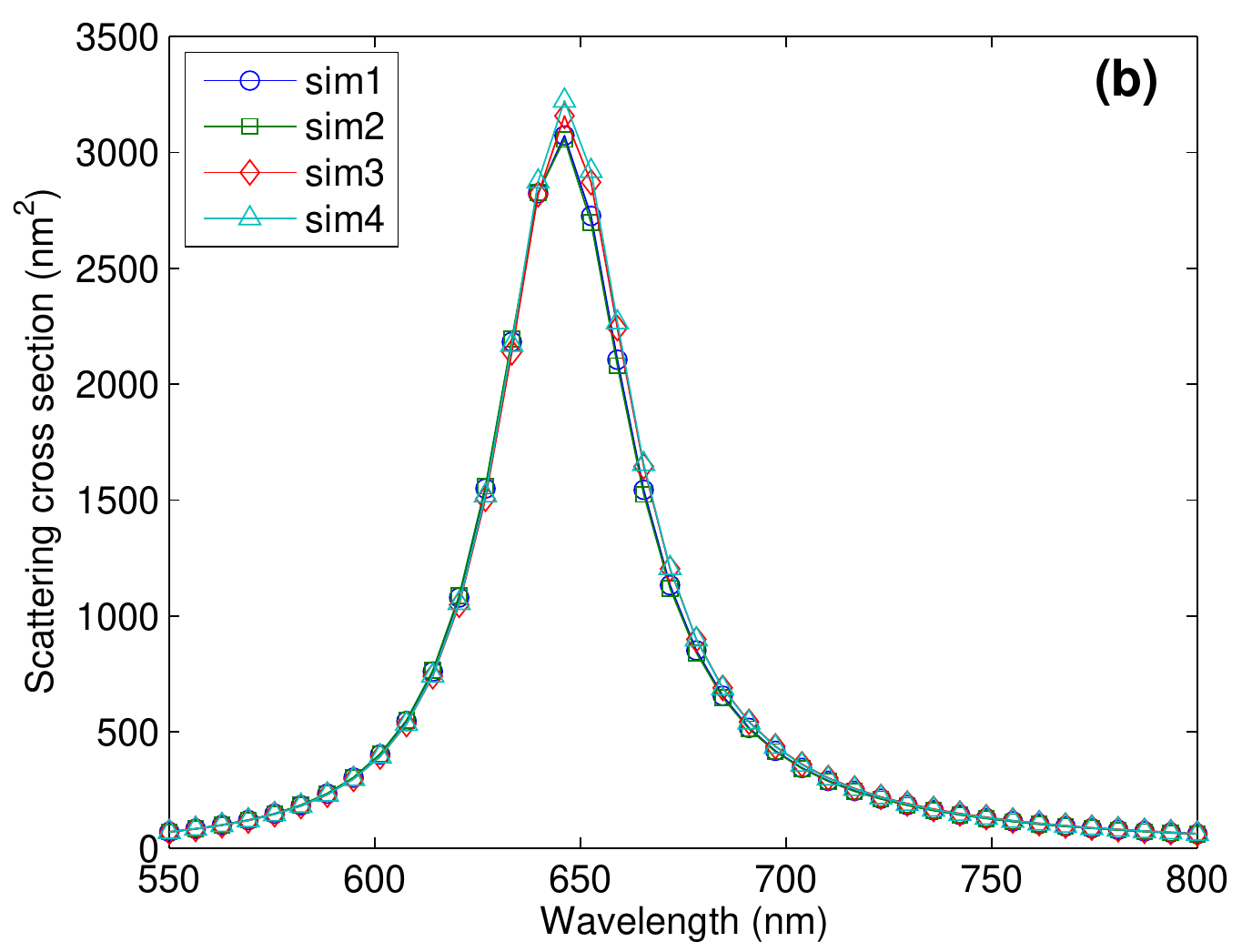}}
\caption{(a) Scattering spectra for plane wave illumination from above (incidence angle of $40^\circ$) and for a gold nanodisk (diameter 60 nm, height 10 nm) that approaches a substrate with a refractive index of 1.5 (glass).  We produce the disk with \texttt{tripolygon(poly,edge)}, where \texttt{poly=polygon(30,'size',[60,60])} and \texttt{edge=edgeprofile(10,11,'mode','01')}.  We also compare the situation where the disk is slightly above the substrate (0.1 nm, refined particle integration with \texttt{refine=3} and \texttt{npol=40}) and the situation where the disk is located on top of the substrate, see text around Eq.~\eqref{eq:singularlayer} for a discussion.  (b) Scattering spectra for disk on substrate and for different simulation scenarios.  \texttt{sim1} is the same simulation as the \texttt{on subs} simulation in panel (a).  In \texttt{sim2} we use through \texttt{edgeprofile(10,25)} a higher number of discretization points in $z$-direction, and in \texttt{sim3} we use through \texttt{polygon(51,'size',[60,60])} a higher number of discretization points for the nanodisk plates.  Finally, in \texttt{sim4} we place the lower disk plate slightly \textit{below} the substrate.  All simulations give very similar results, thus demonstrating the stability and accuracy of our approach.  Computer runtimes are on the order of a few minutes.}\label{fig:disk}
\end{figure}

\subsection{Comparison with DDA}

In Fig.~\ref{fig:dda} we compare results of our BEM implementation with DDA simulations including substrate effects \cite{loke:11} (DDA--SI).  We plot the absorption cross sections for a gold nanosphere (50 nm diameter) located 1 nm above a glass substrate, which is illuminated from above by a plane wave.  The agreement between the two simulations is very good throughout the entire wavelength regime.  In \cite{loke:11} the authors additionally found perfect agreement with complementary FDTD simulations.

\subsection{Comparison with quasistatic simulations}

The MNPBEM toolbox additionally provides the possibility to simulate nanoparticles above substrates within the quasistatic approximation, using the method of image charges for the calculation of the reflected Green function \cite{jackson:99}.  As this approach is completely different from the BEM approach described in the previous sections, a comparison between simulations using the full Maxwell equations (retarded) and simulations employing the quasistatic approximation provide an excellent means for testing our software.

Figure~\ref{fig:stat} presents simulation results for a nanosphere, nanodisk, and nanotriangle situated slightly above a glass substrate.  The nanoparticle dimensions are on the order of a few nanometers, as specified in the figure caption, such that the quasistatic approximation can be safely employed.  Comparison of the full simulation results (solid lines with symbols) with those of the quasistatic approximation (dashed-dotted lines) show excellent agreement for all nanoparticle shapes.

\subsection{Disk above substrate}

We next investigate the situation where a gold nanodisk approaches a glass substrate from above.  In Fig.~\ref{fig:disk}(a) we start with a disk (60 nm diameter, 10 nm height) that is located 10 nm above the substrate, and then gradually decrease the distance to the substrate until the disk is placed on the substrate.  As can be seen, with decreasing distance there is a significant redshift of the plasmon scattering peak.  Most importantly, the simulation results for the disk slightly above (0.1 nm) and directly on the substrate (\texttt{on subs}) give very similar results.  For the particle on the substrate, the lower boundary elements are located less than \verb|layer.ztol| away from the substrate, such that the singular contribution of the reflected Green function can be handled through Eq.~\eqref{eq:singularlayer}.  Here, a few integration points (\texttt{refine=1} and \texttt{npol=10}) suffice to get accurate results.  In contrast, for the 0.1 nm distance (which is larger than \verb|layer.ztol|) we need a much higher number of integration points (\texttt{refine=3} and \texttt{npol=40}) to get converged results.

In Fig.~\ref{fig:disk}(b) we investigate different simulation results for the nanodisk on top of the glass substrate to check the stability and accuracy of our approach.  Increasing the number of $z$-values for the disk (square symbols) or increasing the degree of discretization for the upper and lower disk plates (diamonds, see figure caption for details) give practically indistinguishable results.  We finally performed a simulation where the lower disk plate is placed slightly below the substrate interface.  This can be achieved with
\begin{code}
%  table of dielectric functions
epstab = { epsconst( 1 ), epstable( 'gold.dat' ), epsconst( 2.25 ) };
%  layer structure
layer = layerstructure( epstab, [ 1, 3 ], ztab, op );

%  nanodisk placed slightly above substrate
p1 = tripolygon( polygon( 30, 'size', [ 60, 60 ] ),  ...
             edgeprofile( 10, 11, 'mode', '01', 'min', 1e-3 ) );
%  upper and lower part of nanodisk
[ up, lo ] = select( p1, 'carfun', @( x, y, z ) z > 1.1e-3 );  
%  COMPARTICLE object with plate above substrate
p1 = comparticle( epstab, { p1 }, [ 2, 1 ], 1, op );
%  COMPARTICLE object with plate below substrate
p2 = comparticle( epstab, { up, shift( lo, [ 0, 0, -2e-3 ] }, [ 2, 1; 2, 3 ], [ 1, 2 ], op );
\end{code}
The corresponding simulation results \texttt{sim4} are again in excellent agreement with all previous results.

Quite generally, we found that the particle boundaries must be discretized sufficiently fine around the edges of the lower particle plates in order to get converged results.  This is also evident from the surface charges and currents (not shown) which exhibit strong variations there.  However, from Fig.~\ref{fig:disk} it is obvious that the details of the discretization are not overly critical, and even for coarse discretizations the scattering and extinction spectra are very similar to the converged ones, at least for substrates and layer structures with not too high permittivities.

\begin{figure}
\centerline{\includegraphics[width=0.45\columnwidth]{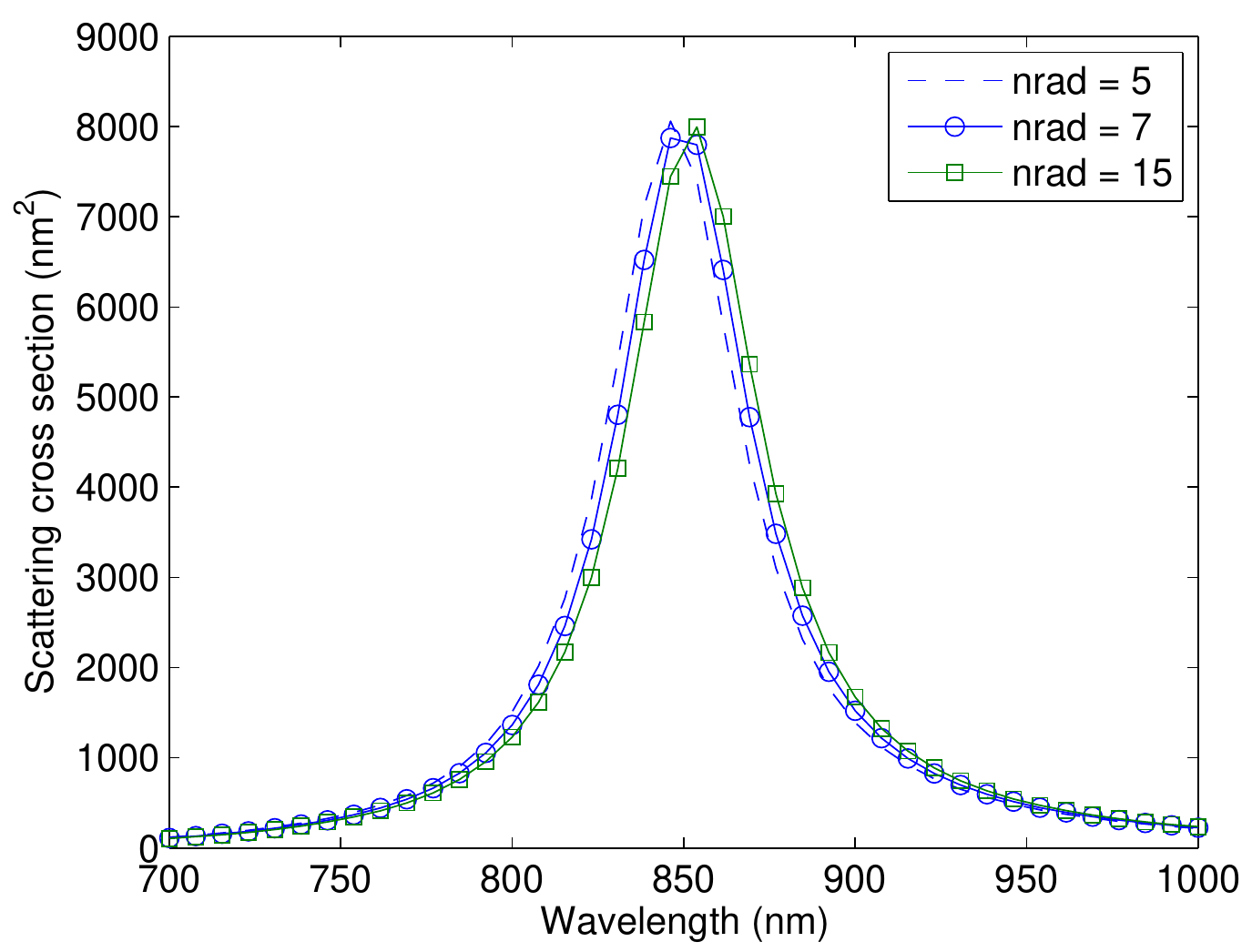}}
\caption{Scattering spectra for gold nanotriangle placed on top of membrane and excited from above with a plane wave.  The nanotriangle has a base length of approximately 80 nm and a height of 10 nm, and is produced with \texttt{tripolygon(poly,edge)} where \texttt{poly=round(polygon(3,'size',[80,80*2/sqrt(3)]),'nrad',nrad)} and \texttt{edge=edgeprofile(10,11,'mode','01')}.  The membrane is a 20 nm thick layer with a dielectric constant of $\varepsilon=4$.  We compare three different discretizations with \texttt{nrad=5,7,15} resulting in 823, 1256 and 2142 boundary elements, respectively.  All discretizations give very similar results.
}\label{fig:tri}
\end{figure}

\subsection{Triangle on membrane}

In Fig.~\ref{fig:tri} we present scattering spectra for a gold nanotriangle placed on top of a membrane with a thickness of 20 nm and a dielectric constant of $\varepsilon=4$.  The dielectric constants of the layer media above and below the membrane are $\varepsilon=1$.  We present three different discretizations, with different numbers of discretization points at the rounded edges of the nanotriangle.  Similar to the previous examples, the influence of the boundary discretization is not very large, although there is a small redshift with increasing discretization indicating that the results are not fully converged.  From additional simulations we observed that the convergence becomes faster when the dielectric constant of the membrane becomes reduced.

\section{Summary}\label{sec:summary}

To summarize, we have developed a methodology for a potential-based boundary element method (BEM) approach to consider substrate and layer structure effects in simulations of plasmonics nanoparticles.  We have implemented our approach within the Matlab MNPBEM toolbox.  A significant speedup of our approach can be achieved by computing the reflected Green functions on a suitable grid and interpolating them at a later stage.  We have compared our implementation with discrete dipole approximation (DDA) simulations including substrate effects as well as with quasistatic simulations, and have found good agreement throughout.  Particular emphasis has been placed on nanoparticles situated directly \textit{on} substrates, as usually encountered in experiment.  We have presented a few representative examples and have demonstrated the viability of our approach.  Typical runtimes for elementary nanoparticle shapes are on the order of a few minutes, although simulations can be sufficiently slower for more complicated nanoparticle geometries.

The new version of the MNPBEM toolbox now includes substrate and layer structure effects, for both retarded (solutions of the full Maxwell equations) and quasistatic simulations.  We have implemented plane wave excitation and the computation of scattering and extinction cross sections, as described in some length in this paper.  Additionally, we provide excitations of oscillating dipoles and the computation of total and radiative scattering rates, which allows one to compute dyadic Green functions and the photonic local density of states.  Such dipole effects have not been discussed in this paper, but some information can be found in the help pages of the toolbox.  Substrate and layer structure effects have not yet been implemented for electron energy loss spectroscopy.

As it stands, the new toolbox provides a flexible toolkit for various plasmonic simulations.  We have tried to implement the new features in a most general fashion, such that a broad class of different situations can be simulated.  So far we have primarily tested substrates and layers consisting of low-permittivity materials.  High-permittivity materials including metals have not been tested properly yet.  They might call for refined particle discretizations, and we thus ask all users to be cautious with the simulation results.  Altogether, we hope that the MNPBEM toolbox will continue to serve the plasmonics community as a useful and helpful simulation software.

\section*{Acknowledgment}

This work has been supported by the Austrian Science Fund FWF under project P24511--N26 and the SFB NextLite, and by NAWI Graz.

\appendix

\section{Deriving the BEM working equations}\label{sec:working}

In this appendix we show how to derive the BEM working equations for layer structures.  We first use Eq.~(\ref{eq:continuity4}a) to express $\bm h_2^\|$ in terms of $h_2^\perp$ and $\sigma_2$.  After some rearrangements we obtain
\begin{equation}\label{eq:app1}
  \left(\Sigma_1-\Sigma_2^\|\right)G_2^\|\bm h_2^\|=
  ik\hat{\bm n}^\|(\varepsilon_2-\varepsilon_1)\bigl(G_2^{\sigma\sigma}\sigma_2+G_2^{\sigma h}h_2^\perp\bigr)+
  \bm\alpha^\|-\Sigma_1\bm a^\|+ik\hat{\bm n}^\|\varepsilon_1\varphi\,,
\end{equation}
with $\Sigma=HG^{-1}$.  With this, we can express the fourth term on the left hand side of Eq.~\eqref{eq:continuity5} as
\begin{equation}
  ik\hat{\bm n}^\|\cdot\bigl(\varepsilon_1 G_1\bm h_1^\|-\varepsilon_2 G_2^\|\bm h_2^\|\bigr)=
  \hat{\bm n}^\|\cdot\Gamma\Bigl[ik\hat{\bm n}^\|(\varepsilon_1-\varepsilon_2)
  \left(G_2^{\sigma\sigma}\sigma_2+G_2^{\sigma h}h_2^\perp\right)+\tilde{\bm\alpha}^\|\Bigr]
  +ik\hat{\bm n}^\|\cdot\varepsilon_1\bm a^\|\,,
\end{equation}
where we have introduced the abbreviations
\begin{displaymath}
  \Gamma=ik(\varepsilon_1-\varepsilon_2)\left(\Sigma_1-\Sigma_2^\|\right)^{-1}\,,\quad
  \tilde{\bm\alpha}=\bm\alpha-\Sigma_1\bm a+ik\hat{\bm n}\varepsilon\varphi\,.
\end{displaymath}
The last term on the left hand side of Eq.~\eqref{eq:continuity5} becomes
\begin{displaymath}
  ik\hat n^\perp\Bigl[\varepsilon_1\left(G_2^{h\sigma}\sigma_2+G_2^{hh}h_2^\perp+a^\perp\right)
  -\varepsilon_2\left(G_2^{h\sigma}\sigma_2+G_2^{hh}h_2^\perp\right)\Bigr]=
  ik\hat n^\perp(\varepsilon_1-\varepsilon_2)\left(G_2^{h\sigma}\sigma_2+G_2^{hh}h_2^\perp\right)
  +ik\hat n^\perp a^\perp\,.
\end{displaymath}
Substituting equation \eqref{eq:app1} into Eq.~\eqref{eq:continuity5} then gives
\begin{eqnarray*}
  &&\varepsilon_1\Sigma_1\left(G_2^{\sigma\sigma}\sigma_2+G_2^{\sigma h}h_2^\perp+\varphi\right)
  -\varepsilon_2 H_2^{\sigma\sigma}\sigma_2-\varepsilon_2H_2^{\sigma h}h_2^\perp
  -ik\hat{\bm n}^\|\cdot\Gamma\hat{\bm n}^\|(\varepsilon_1-\varepsilon_2)
  \left(G_2^{\sigma\sigma}\sigma_2+G_2^{\sigma h}h_2^\perp\right)-\\
  &&\qquad ik\hat n^\perp(\varepsilon_1-\varepsilon_2)\left(G_2^{h\sigma}\sigma_2+G_2^{hh}h_2^\perp\right)
  =D^e+ik\hat{\bm n}^\|\cdot\varepsilon_1\bm a^\|+\hat{\bm n}^\|\cdot\Gamma\tilde{\bm \alpha}^\|+
  ik\hat n^\perp a^\perp\,,
\end{eqnarray*}
which finally leads to the first of the two working equations (\ref{eq:working}a).  The second one is obtained by rewriting Eq.~(\ref{eq:continuity4}b)
\begin{equation}
  \Sigma_1\left(G_2^{h\sigma}\sigma_2+G_2^{hh}h_2^\perp+a^\perp\right)-H_2^{hh}h_2^\perp-H_2^{h\sigma}\sigma_2
  -ik\hat n^\perp\Bigl[\varepsilon_1\left(G_2^{\sigma\sigma}\sigma_2+G_2^{\sigma h}h_2^\perp+\varphi\right)
  -\varepsilon_2\left(G_2^{\sigma\sigma}\sigma_2+G_2^{\sigma h}h_2^\perp\right)\Bigr]=\alpha^\perp\,.
\end{equation}
which leads after some rearrangements to Eq.~(\ref{eq:working}b).

\bigskip


\end{document}